\documentclass[manuscript]{acmart}

\usepackage{multirow}

\AtBeginDocument{%
  \providecommand\BibTeX{{%
    \normalfont B\kern-0.5em{\scshape i\kern-0.25em b}\kern-0.8em\TeX}}}

\copyrightyear{2021}
\acmYear{2021}
\setcopyright{rightsretained}
\acmConference[ASSETS '21]{The 23rd International ACM SIGACCESS
Conference on Computers and Accessibility}{October 18--22,
2021}{Virtual Event, USA}
\acmBooktitle{The 23rd International ACM SIGACCESS Conference on
Computers and Accessibility (ASSETS '21), October 18--22, 2021, Virtual
Event, USA}\acmDOI{10.1145/3441852.3471215}
\acmISBN{978-1-4503-8306-6/21/10}

\begin{document}

\title[Exploring Practices and Challenges of Meal Preparation by People with Visual Impairments]{Non-Visual Cooking: Exploring Practices and Challenges of Meal Preparation by People with Visual Impairments}

\author{Franklin Mingzhe Li}
\affiliation{%
    \institution{Carnegie Mellon University}
    \country{United States}
 }
\email{mingzhe2@cs.cmu.edu}

\author{Jamie Dorst}
\affiliation{%
    \institution{Carnegie Mellon University}
    \country{United States}
 }
 \email{jdorst@andrew.cmu.edu}
 
\author{Peter Cederberg}
\affiliation{%
    \institution{Carnegie Mellon University}
    \country{United States}
 }
 \email{pcederbe@andrew.cmu.edu}
 
\author{Patrick Carrington}
\affiliation{%
    \institution{Carnegie Mellon University}
    \country{United States}
 }
 \email{pcarrington@cmu.edu}

\renewcommand{\shortauthors}{Li et al.}

\begin{abstract}
The reliance on vision for tasks related to cooking and eating healthy can present barriers to cooking for oneself and achieving proper nutrition. There has been little research exploring cooking practices and challenges faced by people with visual impairments. We present a content analysis of 122 YouTube videos to highlight the cooking practices of visually impaired people, and we describe detailed practices for 12 different cooking activities (e.g., cutting and chopping, measuring, testing food for doneness). Based on the cooking practices, we also conducted semi-structured interviews with 12 visually impaired people who have cooking experience and show existing challenges, concerns, and risks in cooking (e.g., tracking the status of tasks in progress, verifying whether things are peeled or cleaned thoroughly). We further discuss opportunities to support the current practices and improve the independence of people with visual impairments in cooking (e.g., zero-touch interactions for cooking). Overall, our findings provide guidance for future research exploring various assistive technologies to help people cook without relying on vision.
\end{abstract}

\ccsdesc[500]{Human-centered computing~Empirical studies in accessibility}

\keywords{accessibility, cooking, people with visual impairments, blind, activity of daily living, assistive technology}


\maketitle

\section{Introduction}


Vision is an important sensory modality for humans. Many activities of daily living (ADLs), such as cooking and eating, can be difficult without visual support. Jones et al. \cite{jones2019analysis} revealed that people with visual impairments tend to have poor nutritional status, which is often linked to problems with buying, preparing, and eating healthy food. People with visual impairments may have an aversion to cooking due to difficulty accessing visual information and cues during the cooking process \cite{bilyk2009food,kostyra2017food}. This has resulted in people with visual impairments more frequently eating outside at restaurants or preparing frozen food that may be calorie-rich. According to the aforementioned Canadian study \cite{bilyk2009food}, eight out of nine participants stated they “disliked or hated cooking” because of the time it takes to cook without vision.


Christine Ha, the first blind contestant of MasterChef, won the third season of the show in 2012 and described the importance of cooking to people with visual impairments \cite{Christin65:online,TheBlind26:online}. To assist visually impaired individuals to cook independently, there have also been training guidelines released from blind communities \cite{SafeCook91:online} and cooking related assistive technologies that are commercially available \cite{HowDoesa82:online}. For example, people with visual impairments could use voice commands to set timers or use a speaking kitchen thermometer to check the temperature of a steak. However, little research has explored the practices of people with visual impairments in cooking and how they leverage different assistive devices for cooking. Furthermore, there has been little to guide HCI researchers on what stages or steps in the cooking process may benefit most from support via technology. In our research, we explore the following research questions:

\begin{itemize}
  \item RQ1: What are current cooking approaches and techniques employed by people with visual impairments?
  \item RQ2: What are the key challenges, concerns, and risks encountered while preparing meals?
  \item RQ3: What are potential opportunities for assistive technologies to support people with visual impairments to cook independently?
\end{itemize}

To first understand the current cooking experiences of people with visual impairments (RQ1), we conducted a content analysis of 122 YouTube videos that feature visually impaired individuals preparing meals. We describe 12 different activities essential to cooking that were summarized from the video analysis. Based on the findings from the video analysis, we then conducted semi-structured interviews with 12 visually impaired people who have experience cooking to better understand RQ2 and RQ3. The interview findings further illuminate challenges encountered before, during, and after cooking including: utilizing tools, information access, touching and feeling, safety and consequence, precision and ambiguity, organizing and tracking, item and quality inspection, and collaborative cooking and communication. We then discuss the potential opportunities to support people with visual impairments while cooking (e.g., zero-touch interactions for cooking, status tracking and safety monitoring, and collaborative cooking).


\section{Background and Related Work}

\subsection{Eating and Cooking for People with Visual Impairments}
People with visual impairments' Activities of Daily Living (e.g., eating and mobility) and Instrumental Activities of Daily Living (e.g., preparing and making food) are affected by the loss of vision \cite{bhowmick2017insight}. Jones et al. \cite{jones2019analysis} conducted a survey study with 101 visually impaired people and found 65\% of the participants stated that their visual impairments made cooking difficult. Due to the difficulty of cooking, Bilyk et al. \cite{bilyk2009food} found that people with visual impairments tend to eat outside or prepared food, which affects healthy eating behaviors. To enable efficient preparation of meals, Kostyra et al. \cite{kostyra2017food} further showed that assistive technologies, such as having equipment with a voice editor, devices informing about the cooking process, and sensors supporting pouring fluids, may enable efficient preparation of meals for people with visual impairments. Therefore, it is important to explore existing cooking practices and challenges for people with visual impairments and understand specific cooking processes or steps that certain assistive technologies may help people with visual impairments in cooking.

\subsection{Enabling Technology for People with Visual Impairments}
Cooking usually requires people with visual impairments to interact with different interfaces or devices. The traditional way to enable people with visual impairments to interact with electronic appliances is by adding tactile markers to them. Beyond this traditional method, prior research also explored using computer vision \cite{guo2016vizlens,fusco2014using,morris2006clearspeech,tekin2011real}, voice interactions \cite{abdolrahmani2018siri,branham2019reading}, and 3D printed tactile marking \cite{guo2017facade,he2017tactile} to better support people with visual impairments interacting with different interfaces. For example, VizLens leveraged computer vision and crowdsourcing to enable people with visual impairments to interact with different interfaces, such as a microwave oven \cite{guo2016vizlens}. Guo et al. \cite{guo2017facade} further introduced a crowdsourced fabrication pipeline to help blind people independently make physical interfaces accessible through adding a 3D printed augmentation of tactile buttons overlaying the original panel. 


Beyond making appliance interfaces accessible, prior research has also explored various approaches to improve the accessibility of mobile devices for people with visual impairments which might help with the cooking process, such as gestural interactions (e.g., \cite{kane2008slide,azenkot2012passchords,li2017braillesketch}) and screen readers (e.g., \cite{rodrigues2015getting,Accessib51:online,leporini2012interacting,Getstart6:online}). For example, Talkback \cite{Getstart6:online}, and VoiceOver \cite{Accessib51:online} enable people with visual impairments to explore interface elements on mobile devices through audio feedback. The feasibility of using mobile devices further allows people with visual impairments to interact with other IoT devices \cite{zhou2017iot,saquib2017blindar}. However, it is unknown how people with visual impairments tend to interact with mobile devices during cooking or utilize their mobile devices to interact with different kitchen appliances, and associated challenges and barriers.

\subsection{Technology for Cooking}
In terms of cooking processes, there has been prior research that explored learning procedures of cooking \cite{kato2013interactive} and different cooking techniques \cite{kusu2017calculating,hamada2005cooking}. For example, Kato and Hasegawa \cite{kato2013interactive} introduced an interactive sauteed cooking simulator that could visualize different cooking states (e.g., temperature changes, browning from burns). This system could help users to better manage their cooking skills, such as how to cook medium-rare meat \cite{kato2013interactive}. Kusu et al. \cite{kusu2017calculating} further proposed a method to calculate a cooking recipe's difficulty level during searching and recommend recipes that match the user's cooking skills. Although prior research has explored how to help people with cooking activities, there lacks research and understanding of what cooking-related learning procedures people with visual impairments have adopted and what the existing challenges are during these learning processes. In our work, we showed cooking practices of 12 different cooking procedures through a YouTube video analysis and uncovered eight themes of cooking challenges through interviews with people with visual impairments.

\section{YouTube Video Analysis: Cooking Practices for People with Visual Impairments}

To understand existing cooking practices and potential risks for people with visual impairments, we conducted a YouTube video analysis---searching, filtering, and analyzing YouTube videos related to cooking practices by people with visual impairments---inspired by prior research on leveraging the richness of YouTube video contents to understand accessibility needs \cite{anthony2013analyzing}. Our video analysis consisted of two main steps: 1) searching for YouTube videos related to cooking practices for people with visual impairments; 2) analysis and coding procedures.

\begin{table}[ht]
\caption{Searching Keywords} 
\centering 
\begin{tabular}{|p{8cm}|} 

\hline 
\textbf{Searching Keywords} \\ 
\hline 
Blind Cooking, Blind Person Cooking, Blind Chef, Legally Blind Cooking, Blind Cooking Food, Blind Cooking Dinner,  Blind in the Kitchen, Visually Impaired Cooking, Visually Impaired Person Cooking, Visually Impaired Chef, Visual Impairment Cooking\\
\hline 
\end{tabular}
\label{table:searchterms} 
\end{table}

\subsection{Search Protocol}
In the video searching process, we looked for videos focused on cooking practices for people with visual impairments. To search for relevant videos, three researchers independently combined visual impairment related keywords (e.g., blind, visually impaired, visual impairment) and cooking related keywords (e.g., cook, cooking, chef, kitchen). To come up with these, our researchers first started with basic searches (e.g., blind cooking) and gradually included other keyword combinations from candidate video titles or descriptions. Because each search may generate hundreds of results, we then followed the same approach as Komkaite et al. \cite{komkaite2019underneath} by stopping our search for videos after the whole page of results started to be irrelevant. 

In total, we initially created a video dataset of 136 relevant videos found by March 28th, 2021. We then filtered out videos if: 1) the person who cooked in the video did not have visual impairments; 2) the person only heated frozen food; 3) it has poor audio and video quality or did not show the person cooking; 4) videos were duplicated. We then ended up filtering 14 videos and created the final video dataset of 122 videos (V1 - V122). Among the 122 videos in our dataset, most videos were uploaded in 2020 (35), while others were uploaded in 2021 (34), 2019 (11), 2018 (10), 2013 (9), 2014 (7), 2017 (6), 2016 (5), 2015 (3), and 2012 (2). The average length of videos was 698 seconds (ranging from 77 seconds to 2694 seconds).

\subsection{Video Content Analysis}
To code the videos, three researchers first open-coded \cite{charmaz2006constructing} the videos independently. Then, the coders met and discussed their codes. When there was a conflict, they explained their rationale for their code to each other and discussed to find a resolution. Eventually, they reached a consensus and consolidated the list of codes. Afterward, we performed affinity diagramming \cite{hartson2012ux} to group the codes and identify emerging themes according to the Cooking Guidelines for people with visual impairments \cite{SafeCook91:online}. Overall, we established 12 themes and 37 codes. The following section describes the findings based on the 12 themes.

\begin{table}[ht]
\small
\caption{Cooking activities from YouTube video analysis for people with visual impairments.} 
\centering 

\begin{tabular}{|p{2.5cm}|p{4cm}|} 

\hline 
Category & Activities\\ [0.5ex] 
\hline 
\multirow{5}{*}{Preparation} & General Safe Cooking Practices\\\cline{2-2}
& Cutting and Chopping\\\cline{2-2}
& Measuring\\\cline{2-2}
& Spreading\\\cline{2-2}
& Pouring\\\hline
\multirow{4}{*}{Cooking} & Placing Pans on a Burner\\\cline{2-2}
& Baking\\\cline{2-2}
& Turning Food\\\cline{2-2}
& Testing Food for Doneness\\\hline
\multirow{3}{2cm}{Tools, Environment, and Recognition} & Tools and Small Appliances\\\cline{2-2}
& Knowing the Kitchen Space\\\cline{2-2}
& Tell Things Apart
\\[0.5ex] 
\hline 
\end{tabular}

\label{table:cookingprocedure} 
\end{table}

\subsection{Results: Cooking Practices of People with Visual Impairments}
From the YouTube video analysis, we uncovered unique cooking practices and procedures of different people with visual impairments (Table \ref{table:cookingprocedure}). 

\subsubsection{General Safe Cooking Practices}
\label{General Safe Cooking Practices}
For general safe cooking tips and practices, we found that people with visual impairments usually \textbf{start everything with gentle actions}. This includes using a slow cooker rather than an open grill to avoid fire flare ups (V18), always cooking on low to medium heat to avoid injury (V80), and putting all food and ingredients in a cold pan first before putting the pan on the burner (V4, V6). In V80, the person with visual impairments mentioned in his video:

\begin{quote}
    ``...For people with visual impairments like me, we might have to make different adjustments on the pan with our hands or utensils while cooking, I often start with using my hand to feel the temperature and cook at low to medium temperature on my pan...''
\end{quote}

Beyond starting with gentle actions, we found that people with visual impairments tend to \textbf{use hearing and hand-feeling to compensate for vision needs during cooking}. In terms of leveraging hands for feeling during cooking, we found that people with visual impairments hold their hands over the pan to gauge temperature (e.g., V11, V23) and use hands to guide food onto utensils (V15). For hearing, we found that people with visual impairments leveraged sound to tell if oil is hot or not (V90, V94). For example, one cook with visual impairments put a small droplet of water into a pan of oil to hear if it was hot (V90). Furthermore, we found that it is important for people with visual impairments to \textbf{memorize the environment in the kitchen}, including the kitchen layout (V1, V17, V46) and which knobs correspond to which burner on the stove (V9).

\begin{figure}[h]
\centering
  \includegraphics[width=1\columnwidth]{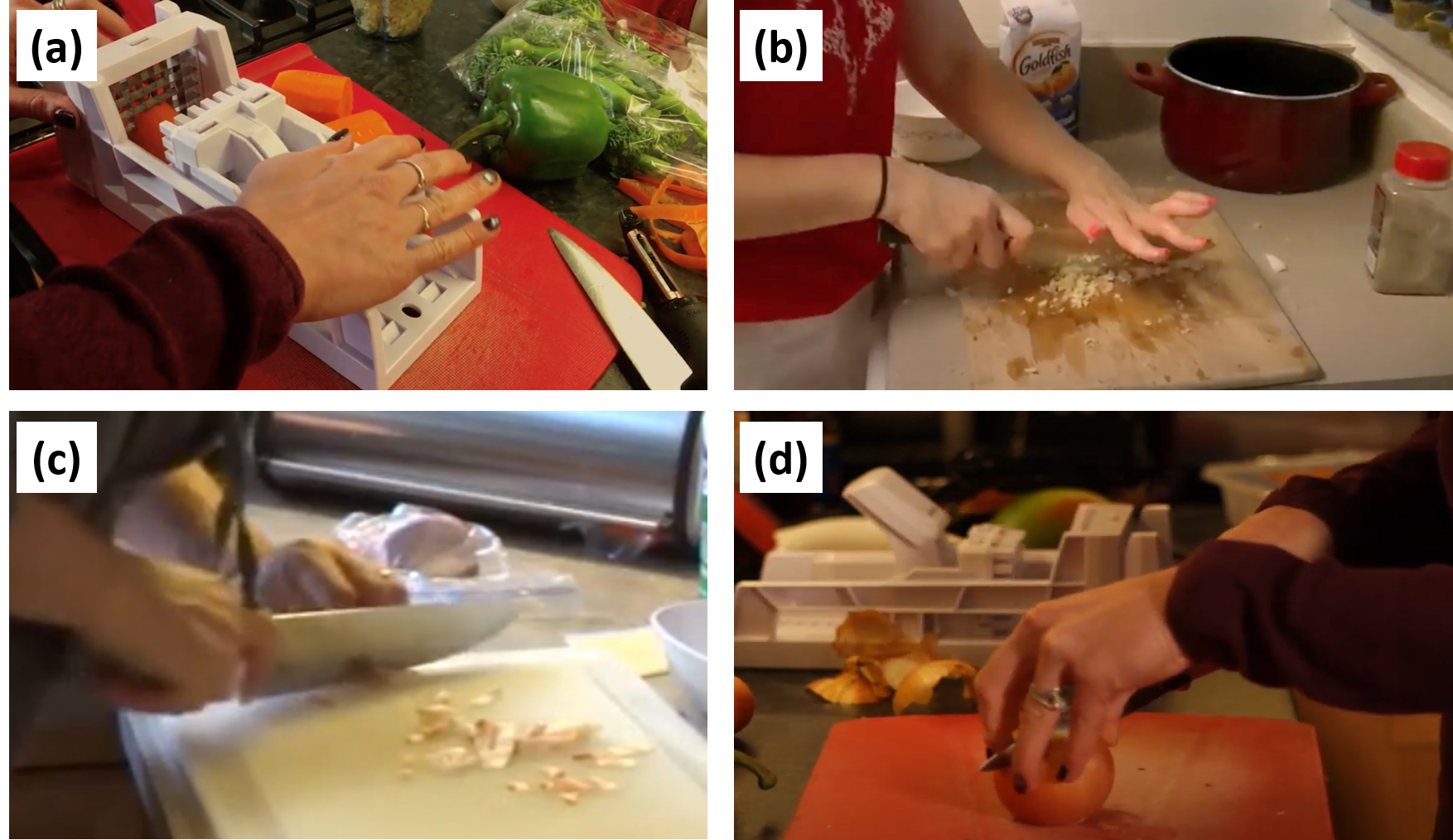}
  \caption{Different ways of cutting. (a) Cutting with a vegetable chopper. (b) Cutting practice of putting the whole hand on top of the knife to cut garlic. (c) Tucking the fingertips under and in towards the palm of the hand and cut with the other hand. (d) Having fingers on both sides of the knife to position where to cut.}~\label{fig:cutting}
  \Description{}
\end{figure}

\subsubsection{Cutting and Chopping}
\label{Cutting and Chopping}
For cutting and chopping, we learned that many people with visual impairments tried \textbf{using alternate ways to avoid knives}, such as using a garlic masher, a vegetable chopper (Fig. \ref{fig:cutting}(a)), a food processor (e.g., V103, V114), scissors (e.g., V84, V94), or even by buying pre-minced garlic (V9, V13). Moreover, we found that people with visual impairments \textbf{sort cut and uncut food into different specific locations on the cutting surface} (V1, V32). In V1, the person commented on her practices in the video: \textit{``For me, I always place the cut food at the top right corner of my cutting board, so I can keep track of the food that I still need to cut.''} We also found that people with visual impairments have \textbf{different practices of using the knife to cut objects}. For example, we found people perform cutting by 1) having fingers on both sides of the knife to position where to cut (e.g., V23, V28) (Fig. \ref{fig:cutting}(d)), 2) putting the whole hand on top of the knife (e.g., V20, V27) (Fig. \ref{fig:cutting}(b)), or 3) tucking the fingertips under and in towards the palm of the hand and cutting with the other hand (e.g., V16, V112) (Fig. \ref{fig:cutting}(c)).

\subsubsection{Measuring}
\label{Measuring}
\begin{figure}[h]
\centering
  \includegraphics[width=0.6\columnwidth]{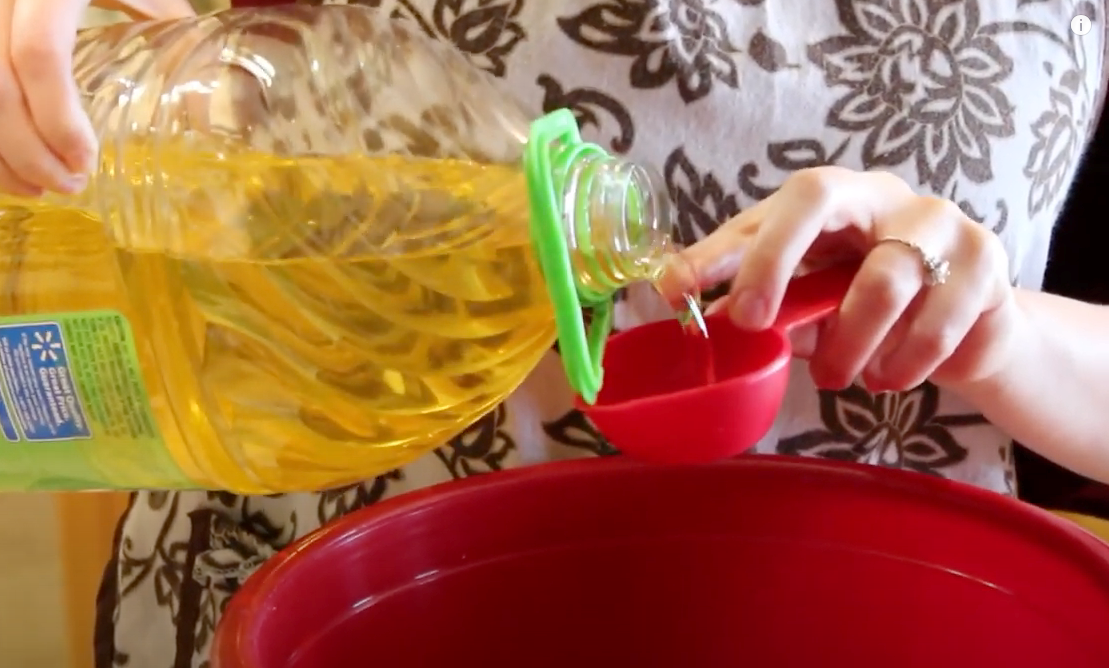}
  \caption{Using the hand to direct the oil and measuring cup to measure liquid (V17).}~\label{fig:measure}
  \Description{}
\end{figure}

To measure food, we first found that there is a common practice for people with visual impairments to \textbf{use hands to weigh meat and measure spices} (e.g., V12, V35). We further revealed that people with visual impairments highly \textbf{rely on using measuring cups with hand assistance to measure liquids} (Fig. \ref{fig:measure}). To accurately use measuring cups, people with visual impairments either have to memorize the measurements (V51) or use a braille-labeled set (V10). Finally, visually impaired cooks also \textbf{prefer using existing containers or utensils to measure food}, such as using the same can to get a 1:1 ratio of water and soup (V89) and using the broad side of a knife to measure the size of meat (V96).

\subsubsection{Pouring}
\label{Pouring}
For pouring, we found that people with visual impairments \textbf{use an extra-large bowl to reduce the difficulty of pouring} (V53, V56) and usually pour over the sink to make sure spills do not matter (V10, V26). To know how much people are pouring, we revealed that people tend to \textbf{use their finger over the spout and/or in containers to know the exact amount} (e.g., V16, V21). Another common practice we discovered is \textbf{\textit{``making sure you poured everything''} to ensure all of the ingredients they want are added to their dish}. More specifically, we found that people first \textbf{add water to food processor or bowl to get any leftover ingredients} (V35) and then \textbf{wipe bowls with fingers to make sure they are empty after pouring} (e.g., V30, V59). 

\subsubsection{Placing Pans on a Burner}
\label{Placing Pans on a Burner}
From the video analysis, we recognized that people with visual impairments \textbf{prefer using traditional electric stove top than modern glass top or gas top}. The key reason was the ease of placing pans on the stove top. As a practice, we found that people with visual impairments tend to either \textbf{use listening or feeling to know whether the pan is on the burner}. For example, V19 showed that visually impaired individuals listen to sizzle to tell whether the pan is on the burner or not after turning on the fire (V19). As another option, we found people either use hand to feel to make sure the pan is at the center of the stove top (V96) or use the back of a spatula to feel if the pan is on the burner (V19).  

\subsubsection{Baking}
\label{Baking}
In baking, we found that visually impaired people have to \textbf{use their hands to feel batter texture, check the batter readiness, and shape dough} (V41, V88). Specifically, we found that visually impaired cooks prefer to fold the dough over parchment paper (V92), use their hands to shape the dough (V121), and create barriers when rolling it out (V82). Furthermore, we revealed \textbf{the importance of keeping things clean during baking}. For example, V29 mentioned the importance of wiping edges of the pan before baking to prevent the burning of anything that accidentally got on the edges. Finally, we found people with visual impairments \textbf{use caution with the oven when baking}, such as by pulling the oven shelf out slightly to avoid burns (V40).

\begin{figure}[h]
\centering
  \includegraphics[width=0.6\columnwidth]{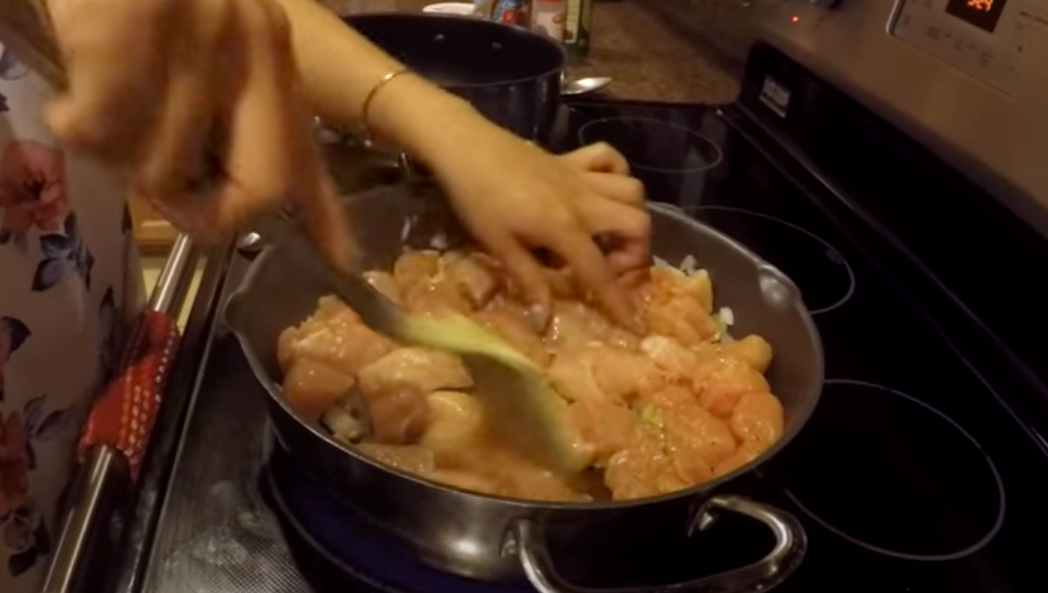}
  \caption{The person uses one hand to feel the food location and another to turn the food with a spatula}~\label{fig:turning}
  \Description{}
\end{figure}

\subsubsection{Turning Food}
\label{Turning Food}
For turning food, we learned that people with visual impairments \textbf{have different preferences on choosing tools} to accomplish the task. This includes using a large two-prong fork (V55), using two utensils (V83), and using tongs and spatulas for (V33, V78). While turning food, people with visual impairments have different practices of \textbf{keeping track of food locations}. For example, V22 showed that the person always memorizes how much food she put on the grill and roughly where it is. V16 further showed that people with visual impairments use one hand to feel for food locations and then use the other hand to turn the food with a spatula (Fig. \ref{fig:turning}).

\subsubsection{Spreading}
\label{Spreading}
From the video analysis, we uncovered that \textbf{using hands to spread} butter or jam is a common practice for people with visual impairments to make sure everything is evenly distributed (e.g., V43, V63). V88 further showed the practices of using single-use plastic gloves for spreading. Beyond using hands, V45 showed the approach of \textbf{using utensils to spread} jam evenly on bread.

\subsubsection{Tools and Small Appliances}
\label{Tools and Small Appliances}
In terms of tools and small appliances for cooking, we found that visually impaired individuals often \textbf{use a virtual assistant to set timers}. This includes using Siri (e.g., V9, V13), Alexa (V1), Google (V4), and analog timers (V44). In V44, the visually impaired person rotated the analog timer using her hand position to gauge the correct time. Furthermore, we uncovered that \textbf{speaking tools are widely used for different purposes}, such as using a barcode reader to know what is in a can (V5), Aira \cite{HomeAira36:online} for identifying cans, screen readers to read recipes off of a phone (V29), video magnifiers to read boxes (V17), talking scales (V41) and speaking thermometers (e.g., V42, V115). We further found that people with visual impairments tend to \textbf{use a lot of specific tools for specific tasks}. For instance, V36 showed that the cook used an automatic can opener and listened for when it was done. Other examples include using a tea strainer to rinse off rice (V84) and spatula and chopsticks for mixing (V95).

\subsubsection{Testing Food for Doneness}
\label{Testing Food for Doneness}
To test whether the food is done, we found that people with visual impairments mostly use five different methods. First, they \textbf{feel to check food doneness}, such as feeling food texture with a spoon (e.g., V9, V86), feeling for fried food to float to the top with a utensil (V90), and feeling cooking food with hands (e.g., V16, V39). We also learned that visually impaired people \textbf{used measures for checking doneness}. For example, they initiated a timer (e.g., V19, V54) or used a thermometer to check whether the food is done (V19, V121). Moreover, we found people with visual impairments also \textbf{used listening to tell when food sounds done} (e.g., V11, V34) and \textbf{use smell to check for food doneness} (e.g., V67, V100). For example, V84 mentioned that visually impaired people could use sound to know whether water is boiling. Finally, we found that people who learned cooking before losing their vision \textbf{judge food doneness on their intuition and prior memories from when they were sighted} (V91).

\subsubsection{Knowing the Kitchen Space}
\label{Knowing the Kitchen Space}
From the YouTube video analysis, we discovered that people with visual impairments put a lot of effort to \textbf{organize the kitchen space to reduce cooking barriers}, such as having separate space for cooking and for ingredients (V10), keeping the container of utensils handle side down so they can feel the ends (V25), keeping measuring cups nested (V28) and always placing utensils back to their original places (V3). Furthermore, we realized visually impaired individuals have different practices to \textbf{feel the kitchen space}. For example, V22 showed that people use a spatula to feel around a hot grill to get to know the space.

\subsubsection{Tell Things Apart}
\label{Tell Things Apart}
To tell ingredients or tools apart, we found a common practice for people with visual impairments is \textbf{adding labels on buttons or objects}. For instance, visually impaired individuals add Braille labels on spice jars (V27, V47) or bump dot stickers for recognizing buttons (V50). Another approach to tell things apart is by \textbf{recognizing the size or shape of containers}. For example, the person from V72 put each ingredient in a different type of bowls to keep track of different ingredients. Furthermore, people also \textbf{rely on smell to tell things apart}, such as smelling to tell the differences between spices (V35, V65). Finally, we found that some people with visual impairments \textbf{prefer putting their containers in certain orders}. For example, some people prefer to alphabetically organize spices in their drawers (V50).

\section{Interview Study: Challenges of Cooking for People with Visual Impairments}
In the YouTube video analysis, we uncovered existing practices for people with visual impairments in different cooking procedures. We then leveraged these findings and conducted semi-structured interviews with visually impaired people who have experience cooking to better understand their existing cooking challenges.

\begin{table}[ht]
\small
\caption{Participants' demographic information} 
\centering 

\begin{tabular}{|p{1.3cm}|p{0.4cm}|p{2cm}|p{4.5cm}|p{3cm}|} 

\hline 
Participant & Age & Gender & Vision Impairment Description & Learned Cooking \textit{Before or After} Vision Loss\\ [0.5ex] 
\hline 
P1 & 25 & Male & Totally Blind, Congenital & After\\
\hline
P2 & 33 & Male & Legally Blind, Congenital & After\\
\hline
P3 & 32 & Non-binary & Legally Blind, Congenital & After\\
\hline
P4 & 19 & Male & Totally Blind, Congenital & After\\
\hline
P5 & 36 & Male & Totally Blind, Acquired (7 years) & Before\\
\hline
P6 & 24 & Female & Legally Blind, Congenital & After\\
\hline
P7 & 22 & Male & Totally Blind, Congenital & After\\
\hline
P8 & 35 & Male & Legally Blind, Acquired (1 year) & Before\\
\hline
P9 & 28 & Female & Totally Blind, Congenital & After\\
\hline
P10 & 31 & Female & Legally Blind, Congenital & After\\
\hline
P11 & 55 & Male & Legally Blind, Congenital & After\\
\hline
P12 & 48 & Female & Legally Blind, Congenital & After \\[0.5ex] 
\hline 
\end{tabular}

\label{table:demographic} 
\end{table}

\subsection{Method: Semi-structured Interview with People with Visual Impairments}
To understand perceptions and challenges of cooking for people with visual impairments, we conducted semi-structured interviews with 12 visually impaired people who have experience cooking (Table \ref{table:demographic}). Our participants have an average age of 32, with a range from 19 to 55 years old. Five of them are totally blind, and the rest are legally blind. Furthermore, ten of our participants are congenitally blind and learned cooking after the vision loss, and the other two cooked before losing their vision. Participants were recruited through social platforms (e.g., Reddit, Twitter, Facebook). To participate in our interview, participants needed to 1) be 18 or above; 2) have visual impairments; 3) have experience cooking; 4) be able to communicate in English. The interviews were conducted through Zoom and took around 60 - 75 minutes for each participant. Participants who completed the interview were compensated by a \$20 Amazon gift card. The entire recruitment and study procedure was approved by the institutional review board (IRB).

In the interview, we first asked participants about their demographic information, general reasoning and perceptions about cooking, and any barriers they encountered when they started cooking. Based on our findings of different cooking procedures and practices in the YouTube video analysis, we asked our participants to discuss their perceptions and challenges around different cooking procedures, such as measuring, pouring, and cutting and chopping. Afterward, we asked participants to discuss their experiences and challenges with assistive cooking technologies, cooking with other people, and the modification of kitchen appliances for accessibility purposes.

Two researchers independently performed open-coding \cite{charmaz2006constructing} on the interview transcripts. Then, the coders met and discussed their codes. When there was a conflict, they explained their rationale for their code to each other and discussed to resolve the conflict. Eventually, they reached a consensus and consolidated the list of codes. Afterward, they performed affinity diagramming \cite{hartson2012ux} to group the codes and identify emerging themes. Overall, we established eight themes and 26 codes. The results introduced in the findings are organized based on eight themes. 

\subsection{Findings}
In this section, we introduce the challenges of cooking derived from our interviews in eight themes: utilizing tools, information access, touching and feeling, safety and consequences, precision and ambiguity, organizing and tracking, item and quality inspection, and collaborative cooking and communication.

\subsubsection{Utilizing Tools}
\label{Utilizing Tools}
From the interviews, we found that people with visual impairments encountered various challenges when utilizing tools (e.g., cooking utensils, appliances). In the video analysis, we found that people with visual impairments rely on either braille markers or bump dots to interact with an interface (Section \ref{Tell Things Apart}). In the interview, seven participants commented on the concerns of \textbf{messing up tools with dirty hands and food during cooking}. For example, P1 commented on the problems of messing up braille tactile markers on kitchen appliances:

\begin{quote}
    ``...It is very common that I might accidentally have something on my hand, especially when I make a dough, then when I touch my Braille markers or bump dots, it just stays on the marker and affects me when I want to read the markers...''
\end{quote}

From the video analysis, we learned that people with visual impairments usually utilize many tools for different purposes (Section \ref{Tools and Small Appliances}). In the interview, we found that eight of our participants \textbf{struggled with having too many tools for different purposes than are necessary}. For example, P12 mentioned that she had to use a vegetable holder on one hand to hold the vegetable and use another hand to peel it. P4 further commented on his situation of even using syringes for cooking:

\begin{quote}
    ``...I have too many tools or cookware for different purposes. I even used syringes to measure and add soy sauce to my dishes. But having too many things really freak me out, it made me look for things all the time...''
\end{quote}

During the interview, five participants mentioned that \textbf{existing tools lack confirmation and feedback}, especially for our participants who learned cooking before vision loss and were used to leverage vision for confirmation and feedback. For example, P8 mentioned the challenges of just using his spatula to tell where food is and whether the food is done: 

\begin{quote}
    ``...Feel through tools is hard, I cannot just use my spatula to do everything for me, it is really hard for me to use it to tell if the food is ready by using tools to feel the texture. That is why I said feel is difficult without hand, that is why we need more help with adding more sensors at the head of tools to tell me different status, and it can also accomplish basic needs...''
\end{quote}

\subsubsection{Information Access}
\label{Information Access}
In the interview, participants overall complained about information access problems with kitchen appliances' guides and recipes. First, three participants mentioned the \textbf{lack of details for guide manuals and instructions of kitchen appliances}. For example, P11 commented on the need for more vocal descriptions about the interface layout: 

\begin{quote}
    ``...I cannot understand existing manuals and instructions for kitchen appliances like pressure cookers, it definitely needs more vocal descriptions about the position of different parts and the interface layout. Especially that the current pressure cookers or rice cookers have so many buttons...''
\end{quote}

Beyond the manuals, we found our participants complained that \textbf{recipe content does not correspond to how people with visual impairments cook}. This includes cookbooks that have too many pictures (P9), wording or cooking language varying among different recipes which confused visually impaired cooks (P7), online recipes containing many figures and ads (P1, P3), and the way current recipes explain different steps requiring lots of vision support (P8). P8 continued with his experiences with recipes like a guessing game:

\begin{quote}
    ``...The descriptions in many recipes are very hard to correspond to the ways that I cook now. I was reading a recipe, and it told me to cook the piece of chicken until it turns brown and then flip it over. It does not transfer to people with visual impairments, it is like a guessing game...''
\end{quote}

Furthermore, we realized that \textbf{existing recipe structures are not friendly to people with visual impairments}. Participants commented that different recipes they found online tend to have different recipe structures, such as the cook's notes, nutrition facts, ingredients and cooking steps. As a consequence, our participants complained that they always have to go back to check the recipe (P1) and do not know where to start the recipe (P1, P3). P6 then further talked about her frustration of starting over from the beginning of the recipe because of the non-standardized recipe structure:

\begin{quote}
    ``...I use voiceover for apple devices, it typically read the whole recipe. However, many recipes do not follow a specific structure which made me easily miss information. Once I lose my place, I have to use voiceover to read from the beginning again...''
\end{quote}

Besides the need for better recipe structures, we also found that our participants \textbf{had a difficult time interacting with the recipe while cooking}. Challenges include problems interacting with phones or tablets while cooking (P1), irrelevant ads and pictures (P10), and worries of getting both physical recipes and electronic devices dirty while cooking (P10). P10 continued: 

\begin{quote}
    ``...I just cannot find a better way to interact with both my braille recipes and electronic recipes by getting them dirty and have a better interactive interface while cooking. Because my hands often have oil or sauce when I am cooking my dishes...''
\end{quote}

Therefore, it is necessary for future research to consider how to better support cooking related information access and better interactive methods while cooking.

\subsubsection{Touching and Feeling}
\label{Touch and Feel}
In the YouTube video analysis, we mentioned that people with visual impairments leverage their hands' feeling for measuring (Section \ref{Measuring}), placing pans on a burner (Section \ref{Placing Pans on a Burner}), baking (Section \ref{Baking}), and testing food for ``doneness" (Section \ref{Testing Food for Doneness}). From the interview, we found that there exist various challenges and concerns about this from people with visual impairments in cooking. First, we found that \textbf{hand feeling is not sufficient to tell all the information about certain cooking procedures}. For example, P12 commented: \textit{``it can be difficult to feel the liquid level if the liquid is room temperature''}. Therefore, P5 complained that he had to put oil in the fridge first and then use a finger to feel the level. P1 further mentioned his difficulties in using his hand to feel the pouring speed of liquid:

\begin{quote}
    ``...When I use my fingers to lead the liquid to pour into a container, I can tell somethings flows over my finger, but I cannot tell how fast that is, and it usually ended up with having too much or too little that got poured, because I do not have another hand to feel the weight of the container, I have to use one hand to pour and use one hand to lead the liquid...''
\end{quote}

Another concern from our participants is that \textbf{using hand feeling during cooking might not be appropriate or unsafe}. As an explanation, P5 mentioned his experiences of messing up the decoration of a cake by using his hands. Moreover, P9 and P12 mentioned that using hands when cooking might cause them to burn themselves by accidentally touching the hot pan. P12 continued:

\begin{quote}
    ``...Using hands to feel the food on a hot pan is not always easy to do, you have to be very very slow and gentle, especially when you want to feel the doneness of the food and try to flip it. I have burned myself many times...''
\end{quote}

Furthermore, we found that visually impaired individuals complained that \textbf{using hands to feel objects affects the efficiency of cooking}, that is, the person cannot use their hand, which is used for feeling and touching, to do other things. As an example we mentioned previously, the person has to use one hand to pour and one hand to lead the liquid to the container, which forces them to place the container at a stable position (P10). Therefore, future research should consider how to enable hand and finger sensing to help with knowing more information during different cooking procedures and explore alternative methods to substitute the need for hand touch and feel in cooking.

\subsubsection{Safety and Consequence}
\label{Safety and Consequence}
From the interview, we found that all participants showed strong concerns of \textbf{getting burned during different cooking procedures}, which need heat detection assistance while cooking. For example, taking things out of the oven puts the visually impaired person in danger (P8), and the plate and dish for baking can be very hot even after taking it out of the oven (P7). P8 explained the danger of exposed heat sources:

\begin{quote}
    ``...A lot of kitchen appliances have exposed heat sources, it can be very dangerous, such as my toaster or the steamer. After I burned myself once, I always pay 200\% attention when I am cooking...''
\end{quote}

Beyond burning, we found participants reported the \textbf{difficulty of handling accidents while cooking}. This includes accidentally knocking things over (P3), it being hard to know if there is an oven fire or not (P10), and accidentally dropping things and making a mess (P5). P5 continued:

\begin{quote}
    ``...I had experiences making scrambled eggs and spilled the egg all over the stovetop. However, I did not notice the mess after cooking, and my partner told me this situation the next day. I need something to alert me or help me to be aware of these things happened...''
\end{quote}

Finally, we found that seven participants mentioned that \textbf{other people walking by could potentially cause safety concerns}, especially for participants who learned cooking before their vision loss. For example, P8 mentioned: \textit{``you may turn around, and someone might stay behind you, and you hold a knife...It is a factor that is out of your control with another person while cooking.''} Thus, future research should explore how to help people with visual impairments detect heat to prevent burns, track accidents, and people walking by.

\subsubsection{Precision and Ambiguity}
\label{Precision and Ambiguity}
In the YouTube video analysis, we introduced the need for measuring (Section \ref{Measuring}) while cooking. From the interview, we found that \textbf{making precise steps while cooking can be a huge barrier for people with visual impairments}, such as getting the right amount of baking soda while baking, because using a hand to feel the level in a measuring cup of powders is challenging (P1). P7 further mentioned the problem with always getting more during spreading:

\begin{quote}
    ``...Because it is hard to know how much butter I already spread on my bread, I always ended up with taking more than needed...''
\end{quote}

P11 further continued with commenting on the difficulty of decorating the food:

\begin{quote}
    ``...Decorating the cake or dishes is extremely hard for me. It took a lot of practice to get used to frosting or adding decorations. And it is very easy to get messed up...''
\end{quote}

To reduce cooking risks, we showed different safe cooking tips in Section \ref{General Safe Cooking Practices}. However, we found that our participants complained about the \textbf{long learning curve of new methods which makes people stick with their familiar way of cooking}. P7 mentioned: \textit{``every person has their own of doing things.''} P3 further commented on the hardness of adopting advice from other people:

\begin{quote}
    ``...It is really difficult to get advice from other people when I am already used to doing things in a certain way. I know my current way of cutting and chopping might not be the recommended way by blind cooking guidelines, but I am already used to it, I tried using the recommended method of cutting and chopping, and I cut myself while learning it...''
\end{quote}

According to participants' responses, another challenge is \textbf{following cooking steps or procedures precisely according to a recipe or instructions}. P10 mentioned that a simple mistake during cooking could end up ruining the dish, such as measuring baking soda and tracking the leavening time during baking. All participants mentioned the correlation between the tedious work of following all cooking steps and procedures and why they do not like to cook. Therefore, future research should explore different tracking and measuring methods to help people with visual impairments follow steps easier and should create various teaching techniques to reduce the learning curve of different safe cooking methods.

\subsubsection{Organizing and Tracking}
\label{Organizing and Tracking}
In Section \ref{Knowing the Kitchen Space}, we showed that people put a lot of effort into organizing their kitchen space to reduce barriers. From the interviews, we uncovered various challenges in organizing the kitchen and tracking objects. First, we found that participants complained about the challenge of \textbf{finding previously used objects or cookware due to the mental load during cooking}. P1 commented: \textit{``It is really easy for me to forget where my used knives are.''} P10 also mentioned that it is hard to find previously used ingredients. P3 then commented on the need of having a system that continuously tracks the kitchen environment:

\begin{quote}
    ``...Cooking itself is a very complicated task that requires people to do multi-tasking and keep tracking of different things. For me, I sometimes forgot where my used knives are, and I had to be very conscious about my actions to prevent being cut by touching my used knife. Therefore, I think I need a tracker that can track the positions of different utensils and cookware for convenience and safety...''
\end{quote}

Furthermore, we found that cooking is not a simple binary task that will guarantee successful and tasty dishes if you follow the recipe. We also realized people with visual impairments \textbf{have a hard time tracking the status of tasks in progress}. Five participants commented that it is hard to know when things are done. They have to constantly check the food. P1 further commented on this problem and tended to overcook the food:

\begin{quote}
    ``...Knowing whether the food is done is such a complicated project for me without any vision. To make sure I do not get diarrhea afterward, I often overcooked the food, such as shrimp. Although I can make sure it is totally cooked, but it affected the texture and the taste of the food...''
\end{quote}

We also found that participants \textbf{have challenges tracking and organizing parts of the food}, such as organizing the cut and uncut parts of a large volume of food, and memorizing which parts of the food on the pan need to be flipped (P10). Therefore, P1 said that she had to wash her hands a lot and touch everything to check the food. P6 continued with her challenges:

\begin{quote}
    ``...The pan is big. If only a portion of the food on the pan needs to be turned over, it would be very difficult. I usually had to use hand and smell to check if the food at the center of the pan is burned or not...''
\end{quote}

Finally, participants mentioned that they often have a hard time \textbf{knowing and remembering where to clean while cooking}. Specifically, six participants complained about remembering where they need to clean afterward, or they have to clean things right after use. P4 further commented that it is also very common for him to not know or not be aware of where he has to clean. Overall, future works should explore how to help people with visual impairments to better organize and track things through cooking. 

\subsubsection{Item and Quality Inspection}
\label{Item and Quality Inspection}
From the interview, we realized that people with visual impairments often have concerns about inspecting an item or food. First of all, we found that participants have a hard time \textbf{checking food quality}, such as whether the bread is moldy. P5 commented on the difficulty of checking food quality just with smell and touch, even with computer vision assistance:

\begin{quote}
    ``...Knowing the food quality is hard just by smelling and touch. Such as knowing whether a piece of bread is moldy or whether bananas or pineapples are ripe. I even tried with some apps that could object recognition, but I realized it just told me the object it is...''
\end{quote}

Furthermore, we found that our participants complained about the task of \textbf{telling whether things got peeled or cleaned thoroughly}. P8 commented on cleaning and peeling non-smooth vegetables:

\begin{quote}
    ``...Different vegetables have different difficulties in cleaning and peeling. Vegetables like carrots are very easy to peel. But non-smooth vegetables are hard to tell if they got peeled or cleaned completely. Ginger is a hard one, also spots on potatoes are hard too...''
\end{quote}

In the interviews, nine participants reported having a hard time \textbf{inspecting an accidental mix of unwanted things}, such as eggshells and bugs in flour (P1). P1 continued: \textit{``I had many times of having eggshells in my scrambled eggs without noticing, it was disgusting.''} Overall, we showed the existing challenges of items and quality inspection. Future research should explore how to bridge the gap of the challenges we introduced to people with visual impairments in cooking.

\subsubsection{Collaborative Cooking and Communication}
\label{Collaborative Cooking and Communication}
From interviewing people with visual impairments, we found that ten out of twelve participants mentioned that they do not like cooking with another person because of \textbf{continuous communication needs}. \textit{``Cooking with another person needs me to constantly speak out things.''} said P4. P3 further complained about collaborative cooking processes:

\begin{quote}
    ``...I have to verbally confirm steps and actions with another person all the time. Sometimes they might just do something in a way that I do not like, and they might not inform me because cooking and talking is not common for any person...''
\end{quote}

In the interview, we found that eight participants leveraged apps for remote assistance while cooking, such as Aira \cite{HomeAira36:online} or Be My Eyes \cite{BeMyEyes54:online}. However, our participants complained about \textbf{making sure the item is in the camera's field of view}. P1 continued:

\begin{quote}
    ``...Remote assistance definitely needs more improvements, I often do not know whether the thing I am talking about is in the camera's field of view. More importantly, my hands are often dirty while cooking, and I always tried not to touch my electronic device. This leads to more concerns about using remote assistance...''
\end{quote}

Moreover, we found that our participants complained about \textbf{appliance modification barriers in a communal space}. P8 commented on this problem: \textit{``the biggest thing about using a shared kitchen, even the simple one like adding bump dots or braille markers, it is hard for me to just add these markers on different appliances, because they are shared.''} Therefore, future research should explore how to improve the social acceptability of adding tactile markers on appliances or other alternative ways for people to interact with different appliances.

\section{Discussion and Future Work}
From the results of our interviews with people with visual impairments, we uncovered existing challenges and gaps for visually impaired people in cooking. In this section, we further discuss zero-touch interactions for cooking, status tracking and safety monitoring, collaborative cooking, and contextual inquiry and learning process in-depth.

\subsection{Zero-touch Interactions for Cooking}
From the results, we showed that our participants often rely on touching and feeling when cooking (e.g., measuring, checking for food doneness, spreading), and interacting with touch-based electronic devices might be inconvenient because they often have food on their hands. As an input method, prior research has explored speech \cite{choi2020nobody, branham2019reading, abdolrahmani2020blind,storer2020all} and mid-air gestures \cite{dim2016designing} as input options for people with visual impairments. However, many of the existing interaction paradigms still require people to physically touch the interfaces (e.g., \cite{ye2014current}), especially to explore elements on the screen. Although visually impaired individuals can currently leverage voice control to input certain commands (setting a timer), future research should explore how to best integrate zero-touch interactions with electronic devices alongside existing practices utilizing touch to manipulate ingredients and cooking tools.

\subsection{Status Tracking and Safety Monitoring}
From our interviews, we highlighted that people with visual impairments have various difficulties in tracking the status of tasks in progress (Section \ref{Organizing and Tracking}), protecting themselves from getting burned during different cooking procedures (Section \ref{Safety and Consequence}), and handling accidents while cooking (Section \ref{Safety and Consequence}). Prior work has explored using thermography to detect heat sources for energy auditing (e.g., \cite{mauriello2017exploring,mauriello2019thermporal}). A similar approach could be applied in the kitchen to detect heat hazards. Beyond heat detection, prior research also explored using computer vision approaches to opportunistically capture actions and provide proactive reminders to users with visual impairments \cite{kianpisheh2019face}. Similar approaches using computer vision or audio \cite{laput2018ubicoustics} could be applied in the kitchen to track object locations or track user's actions and status of cooking steps (e.g., boiling water). 

\subsection{Collaborative Cooking}
The majority of our participants complained about cooking with other people because they have to constantly speak about their actions and ask about other people's actions. Cooking while constantly talking might not be a natural behavior for people both with and without visual impairments. However, three participants did mention the benefit of having another person to accomplish tasks that rely on vision, such as checking expiration dates and food doneness (P10). Prior research has explored collaborative behaviors for people with visual impairments for other purposes, such as gaming \cite{gonccalves2020playing}, use of tangible interfaces \cite{chibaudel2020if}, music composition \cite{omori2013collaborative}, and creating accessible home environments \cite{branham2015collaborative}. Therefore, future research should explore the collaborative cooking behaviors for people with visual impairments and further create guidelines and training methods to reduce safety concerns as well as both mental and physical effort.


\subsection{Contextual Inquiry and Learning Process}
In this paper, we described our research approach using YouTube Video analysis to understand more about the practices of cooking by people with visual impairments. In general, video analyses of this type have the benefit of allowing researchers to gain a broad understanding of the practices of assistive technology use for many different users. This has been exemplified in prior work in HCI and Accessibility (e.g., \cite{anthony2013analyzing,hourcade2015look}). In future research, a contextual inquiry study is well suited to gather a deeper understanding of individual users' practices and challenges (e.g., \cite{dosono2015m,li2019fmt,krome2016contextual,kianpisheh2019face}), which may then be aggregated to understand themes across those individuals. As it relates to the research opportunities we discussed above, a contextual inquiry study may enable future researchers to deeply explore the practices and challenges of zero-touch interactions, status and safety monitoring systems, and collaborative cooking. For example, conducting a contextual inquiry study at a participant's kitchen could identify more status and safety monitoring concerns that might not be recognized by the participants with visual impairments and are not included in the camera's field of view in YouTube videos. An example of this might be watching a visually impaired cook show how she cuts vegetables in front of the camera while, off-camera, cooking chicken in a pan. In this case, neither the audience nor researchers may recognize any potential safety concerns with the pan or other objects in the surroundings from the video. Beyond only interviewing visually impaired individuals, we realized it would also be beneficial for future research to conduct interviews with teachers or trainers who work at community centers and help people with visual impairments with their activities of daily living and meal preparation. This may generate more common practices of cooking by people with visual impairments.


\section{Limitations}
All of the participants in our study were either legally blind or totally blind. We think people with low vision may have different cooking strategies and challenges that utilize their visual perceptive abilities. In our study, we focused on understanding the existing practices and challenges of visually impaired people in their cooking. However, we did also find that people with visual impairments have different learning experiences with cooking. Therefore, people might encounter different challenges at different stages of learning to cook.  

\section{Conclusion}
In this paper, we describe the results of content analyses with 122 YouTube videos featuring visually impaired people preparing meals and uncovered unique cooking practices for people with visual impairments (e.g., use hands to weigh meat and measure spices, use sound and smell to test if the food is done). We then present findings from semi-structured interviews with 12 visually impaired participants and highlight existing challenges encountered while cooking from their perspectives (e.g., tracking the status of tasks in progress, telling whether ingredients are peeled or cleaned thoroughly). We then discussed potential opportunities to integrate technology with the existing practices for cooking, status tracking, safety monitoring, and collaborative cooking. Overall, our findings provide guidance for future research exploring various assistive technologies to help people cook without relying on vision.

\bibliographystyle{ACM-Reference-Format}
\bibliography{main}


\begin{thebibliography}{51}


\ifx \showCODEN    \undefined \def \showCODEN     #1{\unskip}     \fi
\ifx \showDOI      \undefined \def \showDOI       #1{#1}\fi
\ifx \showISBNx    \undefined \def \showISBNx     #1{\unskip}     \fi
\ifx \showISBNxiii \undefined \def \showISBNxiii  #1{\unskip}     \fi
\ifx \showISSN     \undefined \def \showISSN      #1{\unskip}     \fi
\ifx \showLCCN     \undefined \def \showLCCN      #1{\unskip}     \fi
\ifx \shownote     \undefined \def \shownote      #1{#1}          \fi
\ifx \showarticletitle \undefined \def \showarticletitle #1{#1}   \fi
\ifx \showURL      \undefined \def \showURL       {\relax}        \fi
\providecommand\bibfield[2]{#2}
\providecommand\bibinfo[2]{#2}
\providecommand\natexlab[1]{#1}
\providecommand\showeprint[2][]{arXiv:#2}

\bibitem[\protect\citeauthoryear{Abdolrahmani, Kuber, and Branham}{Abdolrahmani
  et~al\mbox{.}}{2018}]%
        {abdolrahmani2018siri}
\bibfield{author}{\bibinfo{person}{Ali Abdolrahmani}, \bibinfo{person}{Ravi
  Kuber}, {and} \bibinfo{person}{Stacy~M Branham}.}
  \bibinfo{year}{2018}\natexlab{}.
\newblock \showarticletitle{" Siri Talks at You" An Empirical Investigation of
  Voice-Activated Personal Assistant (VAPA) Usage by Individuals Who Are
  Blind}. In \bibinfo{booktitle}{\emph{Proceedings of the 20th International
  ACM SIGACCESS Conference on Computers and Accessibility}}.
  \bibinfo{pages}{249--258}.
\newblock


\bibitem[\protect\citeauthoryear{Abdolrahmani, Storer, Roy, Kuber, and
  Branham}{Abdolrahmani et~al\mbox{.}}{2020}]%
        {abdolrahmani2020blind}
\bibfield{author}{\bibinfo{person}{Ali Abdolrahmani}, \bibinfo{person}{Kevin~M
  Storer}, \bibinfo{person}{Antony Rishin~Mukkath Roy}, \bibinfo{person}{Ravi
  Kuber}, {and} \bibinfo{person}{Stacy~M Branham}.}
  \bibinfo{year}{2020}\natexlab{}.
\newblock \showarticletitle{Blind leading the sighted: drawing design insights
  from blind users towards more productivity-oriented voice interfaces}.
\newblock \bibinfo{journal}{\emph{ACM Transactions on Accessible Computing
  (TACCESS)}} \bibinfo{volume}{12}, \bibinfo{number}{4} (\bibinfo{year}{2020}),
  \bibinfo{pages}{1--35}.
\newblock


\bibitem[\protect\citeauthoryear{Anthony, Kim, and Findlater}{Anthony
  et~al\mbox{.}}{2013}]%
        {anthony2013analyzing}
\bibfield{author}{\bibinfo{person}{Lisa Anthony}, \bibinfo{person}{YooJin Kim},
  {and} \bibinfo{person}{Leah Findlater}.} \bibinfo{year}{2013}\natexlab{}.
\newblock \showarticletitle{Analyzing user-generated youtube videos to
  understand touchscreen use by people with motor impairments}. In
  \bibinfo{booktitle}{\emph{Proceedings of the SIGCHI conference on human
  factors in computing systems}}. \bibinfo{pages}{1223--1232}.
\newblock


\bibitem[\protect\citeauthoryear{Azenkot, Rector, Ladner, and Wobbrock}{Azenkot
  et~al\mbox{.}}{2012}]%
        {azenkot2012passchords}
\bibfield{author}{\bibinfo{person}{Shiri Azenkot}, \bibinfo{person}{Kyle
  Rector}, \bibinfo{person}{Richard Ladner}, {and} \bibinfo{person}{Jacob
  Wobbrock}.} \bibinfo{year}{2012}\natexlab{}.
\newblock \showarticletitle{PassChords: secure multi-touch authentication for
  blind people}. In \bibinfo{booktitle}{\emph{Proceedings of the 14th
  international ACM SIGACCESS conference on Computers and accessibility}}.
  \bibinfo{pages}{159--166}.
\newblock


\bibitem[\protect\citeauthoryear{Bhowmick and Hazarika}{Bhowmick and
  Hazarika}{2017}]%
        {bhowmick2017insight}
\bibfield{author}{\bibinfo{person}{Alexy Bhowmick} {and}
  \bibinfo{person}{Shyamanta~M Hazarika}.} \bibinfo{year}{2017}\natexlab{}.
\newblock \showarticletitle{An insight into assistive technology for the
  visually impaired and blind people: state-of-the-art and future trends}.
\newblock \bibinfo{journal}{\emph{Journal on Multimodal User Interfaces}}
  \bibinfo{volume}{11}, \bibinfo{number}{2} (\bibinfo{year}{2017}),
  \bibinfo{pages}{149--172}.
\newblock


\bibitem[\protect\citeauthoryear{Bilyk, Sontrop, Chapman, Barr, and
  Mamer}{Bilyk et~al\mbox{.}}{2009}]%
        {bilyk2009food}
\bibfield{author}{\bibinfo{person}{Marie~Claire Bilyk},
  \bibinfo{person}{Jessica~M Sontrop}, \bibinfo{person}{Gwen~E Chapman},
  \bibinfo{person}{Susan~I Barr}, {and} \bibinfo{person}{Linda Mamer}.}
  \bibinfo{year}{2009}\natexlab{}.
\newblock \showarticletitle{Food experiences and eating patterns of visually
  impaired and blind people}.
\newblock \bibinfo{journal}{\emph{Canadian Journal of Dietetic practice and
  research}} \bibinfo{volume}{70}, \bibinfo{number}{1} (\bibinfo{year}{2009}),
  \bibinfo{pages}{13--18}.
\newblock


\bibitem[\protect\citeauthoryear{Branham and Kane}{Branham and Kane}{2015}]%
        {branham2015collaborative}
\bibfield{author}{\bibinfo{person}{Stacy~M Branham} {and}
  \bibinfo{person}{Shaun~K Kane}.} \bibinfo{year}{2015}\natexlab{}.
\newblock \showarticletitle{Collaborative accessibility: How blind and sighted
  companions co-create accessible home spaces}. In
  \bibinfo{booktitle}{\emph{Proceedings of the 33rd Annual ACM Conference on
  Human Factors in Computing Systems}}. \bibinfo{pages}{2373--2382}.
\newblock


\bibitem[\protect\citeauthoryear{Branham and Mukkath~Roy}{Branham and
  Mukkath~Roy}{2019}]%
        {branham2019reading}
\bibfield{author}{\bibinfo{person}{Stacy~M Branham} {and}
  \bibinfo{person}{Antony~Rishin Mukkath~Roy}.}
  \bibinfo{year}{2019}\natexlab{}.
\newblock \showarticletitle{Reading between the guidelines: How commercial
  voice assistant guidelines hinder accessibility for blind users}. In
  \bibinfo{booktitle}{\emph{The 21st International ACM SIGACCESS Conference on
  Computers and Accessibility}}. \bibinfo{pages}{446--458}.
\newblock


\bibitem[\protect\citeauthoryear{Charmaz}{Charmaz}{2006}]%
        {charmaz2006constructing}
\bibfield{author}{\bibinfo{person}{Kathy Charmaz}.}
  \bibinfo{year}{2006}\natexlab{}.
\newblock \bibinfo{booktitle}{\emph{Constructing grounded theory: A practical
  guide through qualitative analysis}}.
\newblock \bibinfo{publisher}{sage}.
\newblock


\bibitem[\protect\citeauthoryear{Chibaudel, Johal, Oriola, JM~Mac{\'e},
  Dillenbourg, Tartas, and Jouffrais}{Chibaudel et~al\mbox{.}}{2020}]%
        {chibaudel2020if}
\bibfield{author}{\bibinfo{person}{Quentin Chibaudel}, \bibinfo{person}{Wafa
  Johal}, \bibinfo{person}{Bernard Oriola}, \bibinfo{person}{Marc JM~Mac{\'e}},
  \bibinfo{person}{Pierre Dillenbourg}, \bibinfo{person}{Val{\'e}rie Tartas},
  {and} \bibinfo{person}{Christophe Jouffrais}.}
  \bibinfo{year}{2020}\natexlab{}.
\newblock \showarticletitle{" If you've gone straight, now, you must turn
  left"-Exploring the use of a tangible interface in a collaborative treasure
  hunt for people with visual impairments}. In \bibinfo{booktitle}{\emph{The
  22nd International ACM SIGACCESS Conference on Computers and Accessibility}}.
  \bibinfo{pages}{1--10}.
\newblock


\bibitem[\protect\citeauthoryear{Choi, Kwak, Cho, and Lee}{Choi
  et~al\mbox{.}}{2020}]%
        {choi2020nobody}
\bibfield{author}{\bibinfo{person}{Dasom Choi}, \bibinfo{person}{Daehyun Kwak},
  \bibinfo{person}{Minji Cho}, {and} \bibinfo{person}{Sangsu Lee}.}
  \bibinfo{year}{2020}\natexlab{}.
\newblock \showarticletitle{" Nobody Speaks that Fast!" An Empirical Study of
  Speech Rate in Conversational Agents for People with Vision Impairments}. In
  \bibinfo{booktitle}{\emph{Proceedings of the 2020 CHI Conference on Human
  Factors in Computing Systems}}. \bibinfo{pages}{1--13}.
\newblock


\bibitem[\protect\citeauthoryear{Corp}{Corp}{2021}]%
        {HomeAira36:online}
\bibfield{author}{\bibinfo{person}{Aira~Tech Corp}.}
  \bibinfo{year}{2021}\natexlab{}.
\newblock \bibinfo{title}{Home - Aira : Aira}.
\newblock \bibinfo{howpublished}{\url{https://aira.io/}}.
\newblock
\newblock
\shownote{(Accessed on 04/08/2021).}


\bibitem[\protect\citeauthoryear{Dim, Silpasuwanchai, Sarcar, and Ren}{Dim
  et~al\mbox{.}}{2016}]%
        {dim2016designing}
\bibfield{author}{\bibinfo{person}{Nem~Khan Dim}, \bibinfo{person}{Chaklam
  Silpasuwanchai}, \bibinfo{person}{Sayan Sarcar}, {and}
  \bibinfo{person}{Xiangshi Ren}.} \bibinfo{year}{2016}\natexlab{}.
\newblock \showarticletitle{Designing mid-air TV gestures for blind people
  using user-and choice-based elicitation approaches}. In
  \bibinfo{booktitle}{\emph{Proceedings of the 2016 ACM Conference on Designing
  Interactive Systems}}. \bibinfo{pages}{204--214}.
\newblock


\bibitem[\protect\citeauthoryear{Dosono, Hayes, and Wang}{Dosono
  et~al\mbox{.}}{2015}]%
        {dosono2015m}
\bibfield{author}{\bibinfo{person}{Bryan Dosono}, \bibinfo{person}{Jordan
  Hayes}, {and} \bibinfo{person}{Yang Wang}.} \bibinfo{year}{2015}\natexlab{}.
\newblock \showarticletitle{“I’m Stuck!”: A Contextual Inquiry of People
  with Visual Impairments in Authentication}. In
  \bibinfo{booktitle}{\emph{Eleventh Symposium On Usable Privacy and Security
  ($\{$SOUPS$\}$ 2015)}}. \bibinfo{pages}{151--168}.
\newblock


\bibitem[\protect\citeauthoryear{Fusco, Tekin, Ladner, and Coughlan}{Fusco
  et~al\mbox{.}}{2014}]%
        {fusco2014using}
\bibfield{author}{\bibinfo{person}{Giovanni Fusco}, \bibinfo{person}{Ender
  Tekin}, \bibinfo{person}{Richard~E Ladner}, {and} \bibinfo{person}{James~M
  Coughlan}.} \bibinfo{year}{2014}\natexlab{}.
\newblock \showarticletitle{Using computer vision to access appliance
  displays}. In \bibinfo{booktitle}{\emph{Proceedings of the 16th international
  ACM SIGACCESS conference on Computers \& accessibility}}.
  \bibinfo{pages}{281--282}.
\newblock


\bibitem[\protect\citeauthoryear{Gon{\c{c}}alves, Rodrigues, and
  Guerreiro}{Gon{\c{c}}alves et~al\mbox{.}}{2020}]%
        {gonccalves2020playing}
\bibfield{author}{\bibinfo{person}{David Gon{\c{c}}alves},
  \bibinfo{person}{Andr{\'e} Rodrigues}, {and} \bibinfo{person}{Tiago
  Guerreiro}.} \bibinfo{year}{2020}\natexlab{}.
\newblock \showarticletitle{Playing With Others: Depicting Multiplayer Gaming
  Experiences of People With Visual Impairments}. In
  \bibinfo{booktitle}{\emph{The 22nd International ACM SIGACCESS Conference on
  Computers and Accessibility}}. \bibinfo{pages}{1--12}.
\newblock


\bibitem[\protect\citeauthoryear{Guo, Chen, Qi, White, Ghosh, Asakawa, and
  Bigham}{Guo et~al\mbox{.}}{2016}]%
        {guo2016vizlens}
\bibfield{author}{\bibinfo{person}{Anhong Guo}, \bibinfo{person}{Xiang'Anthony'
  Chen}, \bibinfo{person}{Haoran Qi}, \bibinfo{person}{Samuel White},
  \bibinfo{person}{Suman Ghosh}, \bibinfo{person}{Chieko Asakawa}, {and}
  \bibinfo{person}{Jeffrey~P Bigham}.} \bibinfo{year}{2016}\natexlab{}.
\newblock \showarticletitle{Vizlens: A robust and interactive screen reader for
  interfaces in the real world}. In \bibinfo{booktitle}{\emph{Proceedings of
  the 29th Annual Symposium on User Interface Software and Technology}}.
  \bibinfo{pages}{651--664}.
\newblock


\bibitem[\protect\citeauthoryear{Guo, Kim, Chen, Yeh, Hudson, Mankoff, and
  Bigham}{Guo et~al\mbox{.}}{2017}]%
        {guo2017facade}
\bibfield{author}{\bibinfo{person}{Anhong Guo}, \bibinfo{person}{Jeeeun Kim},
  \bibinfo{person}{Xiang'Anthony' Chen}, \bibinfo{person}{Tom Yeh},
  \bibinfo{person}{Scott~E Hudson}, \bibinfo{person}{Jennifer Mankoff}, {and}
  \bibinfo{person}{Jeffrey~P Bigham}.} \bibinfo{year}{2017}\natexlab{}.
\newblock \showarticletitle{Facade: Auto-generating tactile interfaces to
  appliances}. In \bibinfo{booktitle}{\emph{Proceedings of the 2017 CHI
  Conference on Human Factors in Computing Systems}}.
  \bibinfo{pages}{5826--5838}.
\newblock


\bibitem[\protect\citeauthoryear{Ha}{Ha}{2020}]%
        {TheBlind26:online}
\bibfield{author}{\bibinfo{person}{Christine Ha}.}
  \bibinfo{year}{2020}\natexlab{}.
\newblock \bibinfo{title}{The Blind Cook – Christine Ha's Adventures}.
\newblock \bibinfo{howpublished}{\url{http://www.theblindcook.com/}}.
\newblock
\newblock
\shownote{(Accessed on 03/14/2021).}


\bibitem[\protect\citeauthoryear{Hamada, Okabe, Ide, Satoh, Sakai, and
  Tanaka}{Hamada et~al\mbox{.}}{2005}]%
        {hamada2005cooking}
\bibfield{author}{\bibinfo{person}{Reiko Hamada}, \bibinfo{person}{Jun Okabe},
  \bibinfo{person}{Ichiro Ide}, \bibinfo{person}{Shin'ichi Satoh},
  \bibinfo{person}{Shuichi Sakai}, {and} \bibinfo{person}{Hidehiko Tanaka}.}
  \bibinfo{year}{2005}\natexlab{}.
\newblock \showarticletitle{Cooking navi: assistant for daily cooking in
  kitchen}. In \bibinfo{booktitle}{\emph{Proceedings of the 13th annual ACM
  international conference on Multimedia}}. \bibinfo{pages}{371--374}.
\newblock


\bibitem[\protect\citeauthoryear{Hartson and Pyla}{Hartson and Pyla}{2012}]%
        {hartson2012ux}
\bibfield{author}{\bibinfo{person}{Rex Hartson} {and} \bibinfo{person}{Pardha~S
  Pyla}.} \bibinfo{year}{2012}\natexlab{}.
\newblock \bibinfo{booktitle}{\emph{The UX Book: Process and guidelines for
  ensuring a quality user experience}}.
\newblock \bibinfo{publisher}{Elsevier}.
\newblock


\bibitem[\protect\citeauthoryear{He, Wan, Findlater, and Froehlich}{He
  et~al\mbox{.}}{2017}]%
        {he2017tactile}
\bibfield{author}{\bibinfo{person}{Liang He}, \bibinfo{person}{Zijian Wan},
  \bibinfo{person}{Leah Findlater}, {and} \bibinfo{person}{Jon~E Froehlich}.}
  \bibinfo{year}{2017}\natexlab{}.
\newblock \showarticletitle{TacTILE: a preliminary toolchain for creating
  accessible graphics with 3D-printed overlays and auditory annotations}. In
  \bibinfo{booktitle}{\emph{Proceedings of the 19th International ACM SIGACCESS
  Conference on Computers and Accessibility}}. \bibinfo{pages}{397--398}.
\newblock


\bibitem[\protect\citeauthoryear{Henkler}{Henkler}{2020}]%
        {HowDoesa82:online}
\bibfield{author}{\bibinfo{person}{Ed Henkler}.}
  \bibinfo{year}{2020}\natexlab{}.
\newblock \bibinfo{title}{How Does a Blind Person Cook? - The Blind Guide}.
\newblock
  \bibinfo{howpublished}{\url{https://theblindguide.com/how-does-blind-person-cook/}}.
\newblock
\newblock
\shownote{(Accessed on 03/14/2021).}


\bibitem[\protect\citeauthoryear{Hourcade, Mascher, Wu, and Pantoja}{Hourcade
  et~al\mbox{.}}{2015}]%
        {hourcade2015look}
\bibfield{author}{\bibinfo{person}{Juan~Pablo Hourcade},
  \bibinfo{person}{Sarah~L Mascher}, \bibinfo{person}{David Wu}, {and}
  \bibinfo{person}{Luiza Pantoja}.} \bibinfo{year}{2015}\natexlab{}.
\newblock \showarticletitle{Look, my baby is using an iPad! An analysis of
  YouTube videos of infants and toddlers using tablets}. In
  \bibinfo{booktitle}{\emph{Proceedings of the 33rd Annual ACM Conference on
  Human Factors in Computing Systems}}. \bibinfo{pages}{1915--1924}.
\newblock


\bibitem[\protect\citeauthoryear{Inc.}{Inc.}{2021}]%
        {Accessib51:online}
\bibfield{author}{\bibinfo{person}{Apple Inc.}}
  \bibinfo{year}{2021}\natexlab{}.
\newblock \bibinfo{title}{Accessibility - Vision - Apple (CA)}.
\newblock
  \bibinfo{howpublished}{\url{https://www.apple.com/ca/accessibility/vision/}}.
\newblock
\newblock
\shownote{(Accessed on 03/16/2021).}


\bibitem[\protect\citeauthoryear{Jones, Bartlett, and Cooke}{Jones
  et~al\mbox{.}}{2019}]%
        {jones2019analysis}
\bibfield{author}{\bibinfo{person}{Nabila Jones},
  \bibinfo{person}{Hannah~Elizabeth Bartlett}, {and} \bibinfo{person}{Richard
  Cooke}.} \bibinfo{year}{2019}\natexlab{}.
\newblock \showarticletitle{An analysis of the impact of visual impairment on
  activities of daily living and vision-related quality of life in a visually
  impaired adult population}.
\newblock \bibinfo{journal}{\emph{British Journal of Visual Impairment}}
  \bibinfo{volume}{37}, \bibinfo{number}{1} (\bibinfo{year}{2019}),
  \bibinfo{pages}{50--63}.
\newblock


\bibitem[\protect\citeauthoryear{Kane, Bigham, and Wobbrock}{Kane
  et~al\mbox{.}}{2008}]%
        {kane2008slide}
\bibfield{author}{\bibinfo{person}{Shaun~K Kane}, \bibinfo{person}{Jeffrey~P
  Bigham}, {and} \bibinfo{person}{Jacob~O Wobbrock}.}
  \bibinfo{year}{2008}\natexlab{}.
\newblock \showarticletitle{Slide rule: making mobile touch screens accessible
  to blind people using multi-touch interaction techniques}. In
  \bibinfo{booktitle}{\emph{Proceedings of the 10th international ACM SIGACCESS
  conference on Computers and accessibility}}. \bibinfo{pages}{73--80}.
\newblock


\bibitem[\protect\citeauthoryear{Kato and Hasegawa}{Kato and Hasegawa}{2013}]%
        {kato2013interactive}
\bibfield{author}{\bibinfo{person}{Fumihiro Kato} {and}
  \bibinfo{person}{Shoichi Hasegawa}.} \bibinfo{year}{2013}\natexlab{}.
\newblock \showarticletitle{Interactive Cooking Simulator: Showing food
  ingredients appearance changes in frying pan cooking}. In
  \bibinfo{booktitle}{\emph{Proceedings of the 5th international workshop on
  Multimedia for cooking \& eating activities}}. \bibinfo{pages}{33--38}.
\newblock


\bibitem[\protect\citeauthoryear{Kianpisheh, Li, and Truong}{Kianpisheh
  et~al\mbox{.}}{2019}]%
        {kianpisheh2019face}
\bibfield{author}{\bibinfo{person}{Mohammad Kianpisheh},
  \bibinfo{person}{Franklin~Mingzhe Li}, {and} \bibinfo{person}{Khai~N
  Truong}.} \bibinfo{year}{2019}\natexlab{}.
\newblock \showarticletitle{Face recognition assistant for people with visual
  impairments}.
\newblock \bibinfo{journal}{\emph{Proceedings of the ACM on Interactive,
  Mobile, Wearable and Ubiquitous Technologies}} \bibinfo{volume}{3},
  \bibinfo{number}{3} (\bibinfo{year}{2019}), \bibinfo{pages}{1--24}.
\newblock


\bibitem[\protect\citeauthoryear{Komkaite, Lavrinovica, Vraka, and
  Skov}{Komkaite et~al\mbox{.}}{2019}]%
        {komkaite2019underneath}
\bibfield{author}{\bibinfo{person}{Aida Komkaite}, \bibinfo{person}{Liga
  Lavrinovica}, \bibinfo{person}{Maria Vraka}, {and} \bibinfo{person}{Mikael~B
  Skov}.} \bibinfo{year}{2019}\natexlab{}.
\newblock \showarticletitle{Underneath the Skin: An Analysis of YouTube Videos
  to Understand Insertable Device Interaction}. In
  \bibinfo{booktitle}{\emph{Proceedings of the 2019 CHI Conference on Human
  Factors in Computing Systems}}. \bibinfo{pages}{1--12}.
\newblock


\bibitem[\protect\citeauthoryear{Kostyra, {\.Z}akowska-Biemans, {\'S}niegocka,
  and Piotrowska}{Kostyra et~al\mbox{.}}{2017}]%
        {kostyra2017food}
\bibfield{author}{\bibinfo{person}{Eliza Kostyra}, \bibinfo{person}{Sylwia
  {\.Z}akowska-Biemans}, \bibinfo{person}{Katarzyna {\'S}niegocka}, {and}
  \bibinfo{person}{Anna Piotrowska}.} \bibinfo{year}{2017}\natexlab{}.
\newblock \showarticletitle{Food shopping, sensory determinants of food choice
  and meal preparation by visually impaired people. Obstacles and expectations
  in daily food experiences}.
\newblock \bibinfo{journal}{\emph{Appetite}}  \bibinfo{volume}{113}
  (\bibinfo{year}{2017}), \bibinfo{pages}{14--22}.
\newblock


\bibitem[\protect\citeauthoryear{Krome, Walz, and Greuter}{Krome
  et~al\mbox{.}}{2016}]%
        {krome2016contextual}
\bibfield{author}{\bibinfo{person}{Sven Krome}, \bibinfo{person}{Steffen~P
  Walz}, {and} \bibinfo{person}{Stefan Greuter}.}
  \bibinfo{year}{2016}\natexlab{}.
\newblock \showarticletitle{Contextual inquiry of future commuting in
  autonomous cars}. In \bibinfo{booktitle}{\emph{Proceedings of the 2016 CHI
  Conference Extended Abstracts on Human Factors in Computing Systems}}.
  \bibinfo{pages}{3122--3128}.
\newblock


\bibitem[\protect\citeauthoryear{Kusu, Makino, Shioi, and Hatano}{Kusu
  et~al\mbox{.}}{2017}]%
        {kusu2017calculating}
\bibfield{author}{\bibinfo{person}{Kazuma Kusu}, \bibinfo{person}{Nozomi
  Makino}, \bibinfo{person}{Takamitsu Shioi}, {and} \bibinfo{person}{Kenji
  Hatano}.} \bibinfo{year}{2017}\natexlab{}.
\newblock \showarticletitle{Calculating Cooking Recipe's Difficulty based on
  Cooking Activities}. In \bibinfo{booktitle}{\emph{Proceedings of the 9th
  Workshop on Multimedia for Cooking and Eating Activities in conjunction with
  The 2017 International Joint Conference on Artificial Intelligence}}.
  \bibinfo{pages}{19--24}.
\newblock


\bibitem[\protect\citeauthoryear{Laput, Ahuja, Goel, and Harrison}{Laput
  et~al\mbox{.}}{2018}]%
        {laput2018ubicoustics}
\bibfield{author}{\bibinfo{person}{Gierad Laput}, \bibinfo{person}{Karan
  Ahuja}, \bibinfo{person}{Mayank Goel}, {and} \bibinfo{person}{Chris
  Harrison}.} \bibinfo{year}{2018}\natexlab{}.
\newblock \showarticletitle{Ubicoustics: Plug-and-play acoustic activity
  recognition}. In \bibinfo{booktitle}{\emph{Proceedings of the 31st Annual ACM
  Symposium on User Interface Software and Technology}}.
  \bibinfo{pages}{213--224}.
\newblock


\bibitem[\protect\citeauthoryear{Leporini, Buzzi, and Buzzi}{Leporini
  et~al\mbox{.}}{2012}]%
        {leporini2012interacting}
\bibfield{author}{\bibinfo{person}{Barbara Leporini},
  \bibinfo{person}{Maria~Claudia Buzzi}, {and} \bibinfo{person}{Marina Buzzi}.}
  \bibinfo{year}{2012}\natexlab{}.
\newblock \showarticletitle{Interacting with mobile devices via VoiceOver:
  usability and accessibility issues}. In \bibinfo{booktitle}{\emph{Proceedings
  of the 24th Australian Computer-Human Interaction Conference}}.
  \bibinfo{pages}{339--348}.
\newblock


\bibitem[\protect\citeauthoryear{Li, Chen, Fan, and Truong}{Li
  et~al\mbox{.}}{2019}]%
        {li2019fmt}
\bibfield{author}{\bibinfo{person}{Franklin~Mingzhe Li},
  \bibinfo{person}{Di~Laura Chen}, \bibinfo{person}{Mingming Fan}, {and}
  \bibinfo{person}{Khai~N Truong}.} \bibinfo{year}{2019}\natexlab{}.
\newblock \showarticletitle{FMT: A Wearable Camera-Based Object Tracking Memory
  Aid for Older Adults}.
\newblock \bibinfo{journal}{\emph{Proceedings of the ACM on Interactive,
  Mobile, Wearable and Ubiquitous Technologies}} \bibinfo{volume}{3},
  \bibinfo{number}{3} (\bibinfo{year}{2019}), \bibinfo{pages}{1--25}.
\newblock


\bibitem[\protect\citeauthoryear{Li, Fan, and Truong}{Li et~al\mbox{.}}{2017}]%
        {li2017braillesketch}
\bibfield{author}{\bibinfo{person}{Mingzhe Li}, \bibinfo{person}{Mingming Fan},
  {and} \bibinfo{person}{Khai~N Truong}.} \bibinfo{year}{2017}\natexlab{}.
\newblock \showarticletitle{BrailleSketch: A gesture-based text input method
  for people with visual impairments}. In \bibinfo{booktitle}{\emph{Proceedings
  of the 19th International ACM SIGACCESS Conference on Computers and
  Accessibility}}. \bibinfo{pages}{12--21}.
\newblock


\bibitem[\protect\citeauthoryear{LLC}{LLC}{2021}]%
        {Getstart6:online}
\bibfield{author}{\bibinfo{person}{Google LLC}.}
  \bibinfo{year}{2021}\natexlab{}.
\newblock \bibinfo{title}{Get started on Android with TalkBack - Android
  Accessibility Help}.
\newblock
  \bibinfo{howpublished}{\url{https://support.google.com/accessibility/android/answer/6283677?hl=en}}.
\newblock
\newblock
\shownote{(Accessed on 03/16/2021).}


\bibitem[\protect\citeauthoryear{Mauriello, McNally, and Froehlich}{Mauriello
  et~al\mbox{.}}{2019}]%
        {mauriello2019thermporal}
\bibfield{author}{\bibinfo{person}{Matthew~Louis Mauriello},
  \bibinfo{person}{Brenna McNally}, {and} \bibinfo{person}{Jon~E Froehlich}.}
  \bibinfo{year}{2019}\natexlab{}.
\newblock \showarticletitle{Thermporal: An easy-to-deploy temporal
  thermographic sensor system to support residential energy audits}. In
  \bibinfo{booktitle}{\emph{Proceedings of the 2019 CHI Conference on Human
  Factors in Computing Systems}}. \bibinfo{pages}{1--14}.
\newblock


\bibitem[\protect\citeauthoryear{Mauriello, Saha, Brown, and
  Froehlich}{Mauriello et~al\mbox{.}}{2017}]%
        {mauriello2017exploring}
\bibfield{author}{\bibinfo{person}{Matthew~Louis Mauriello},
  \bibinfo{person}{Manaswi Saha}, \bibinfo{person}{Erica~Brown Brown}, {and}
  \bibinfo{person}{Jon~E Froehlich}.} \bibinfo{year}{2017}\natexlab{}.
\newblock \showarticletitle{Exploring novice approaches to smartphone-based
  thermographic energy auditing: A field study}. In
  \bibinfo{booktitle}{\emph{Proceedings of the 2017 CHI Conference on Human
  Factors in Computing Systems}}. \bibinfo{pages}{1768--1780}.
\newblock


\bibitem[\protect\citeauthoryear{Morris, Blenkhorn, Crossey, Ngo, Ross, Werner,
  and Wong}{Morris et~al\mbox{.}}{2006}]%
        {morris2006clearspeech}
\bibfield{author}{\bibinfo{person}{Tim Morris}, \bibinfo{person}{Paul
  Blenkhorn}, \bibinfo{person}{Luke Crossey}, \bibinfo{person}{Quang Ngo},
  \bibinfo{person}{Martin Ross}, \bibinfo{person}{David Werner}, {and}
  \bibinfo{person}{Christina Wong}.} \bibinfo{year}{2006}\natexlab{}.
\newblock \showarticletitle{Clearspeech: A display reader for the visually
  handicapped}.
\newblock \bibinfo{journal}{\emph{IEEE Transactions on Neural Systems and
  Rehabilitation Engineering}} \bibinfo{volume}{14}, \bibinfo{number}{4}
  (\bibinfo{year}{2006}), \bibinfo{pages}{492--500}.
\newblock


\bibitem[\protect\citeauthoryear{Omori and Yairi}{Omori and Yairi}{2013}]%
        {omori2013collaborative}
\bibfield{author}{\bibinfo{person}{Shotaro Omori} {and}
  \bibinfo{person}{Ikuko~Eguchi Yairi}.} \bibinfo{year}{2013}\natexlab{}.
\newblock \showarticletitle{Collaborative music application for visually
  impaired people with tangible objects on table}. In
  \bibinfo{booktitle}{\emph{Proceedings of the 15th International ACM SIGACCESS
  Conference on Computers and Accessibility}}. \bibinfo{pages}{1--2}.
\newblock


\bibitem[\protect\citeauthoryear{Rodrigues, Montague, Nicolau, and
  Guerreiro}{Rodrigues et~al\mbox{.}}{2015}]%
        {rodrigues2015getting}
\bibfield{author}{\bibinfo{person}{Andr{\'e} Rodrigues}, \bibinfo{person}{Kyle
  Montague}, \bibinfo{person}{Hugo Nicolau}, {and} \bibinfo{person}{Tiago
  Guerreiro}.} \bibinfo{year}{2015}\natexlab{}.
\newblock \showarticletitle{Getting smartphones to talkback: Understanding the
  smartphone adoption process of blind users}. In
  \bibinfo{booktitle}{\emph{Proceedings of the 17th international acm sigaccess
  conference on computers \& accessibility}}. \bibinfo{pages}{23--32}.
\newblock


\bibitem[\protect\citeauthoryear{Saquib, Murari, and Bhargav}{Saquib
  et~al\mbox{.}}{2017}]%
        {saquib2017blindar}
\bibfield{author}{\bibinfo{person}{Zeeshan Saquib}, \bibinfo{person}{Vishakha
  Murari}, {and} \bibinfo{person}{Suhas~N Bhargav}.}
  \bibinfo{year}{2017}\natexlab{}.
\newblock \showarticletitle{BlinDar: An invisible eye for the blind people
  making life easy for the blind with Internet of Things (IoT)}. In
  \bibinfo{booktitle}{\emph{2017 2nd IEEE International Conference on Recent
  Trends in Electronics, Information \& Communication Technology (RTEICT)}}.
  IEEE, \bibinfo{pages}{71--75}.
\newblock


\bibitem[\protect\citeauthoryear{Storer, Judge, and Branham}{Storer
  et~al\mbox{.}}{2020}]%
        {storer2020all}
\bibfield{author}{\bibinfo{person}{Kevin~M Storer}, \bibinfo{person}{Tejinder~K
  Judge}, {and} \bibinfo{person}{Stacy~M Branham}.}
  \bibinfo{year}{2020}\natexlab{}.
\newblock \showarticletitle{" All in the Same Boat": Tradeoffs of Voice
  Assistant Ownership for Mixed-Visual-Ability Families}. In
  \bibinfo{booktitle}{\emph{Proceedings of the 2020 CHI Conference on Human
  Factors in Computing Systems}}. \bibinfo{pages}{1--14}.
\newblock


\bibitem[\protect\citeauthoryear{Tekin, Coughlan, and Shen}{Tekin
  et~al\mbox{.}}{2011}]%
        {tekin2011real}
\bibfield{author}{\bibinfo{person}{Ender Tekin}, \bibinfo{person}{James~M
  Coughlan}, {and} \bibinfo{person}{Huiying Shen}.}
  \bibinfo{year}{2011}\natexlab{}.
\newblock \showarticletitle{Real-time detection and reading of LED/LCD displays
  for visually impaired persons}. In \bibinfo{booktitle}{\emph{2011 IEEE
  Workshop on Applications of Computer Vision (WACV)}}. IEEE,
  \bibinfo{pages}{491--496}.
\newblock


\bibitem[\protect\citeauthoryear{VisionAware}{VisionAware}{[n.d.]}]%
        {SafeCook91:online}
\bibfield{author}{\bibinfo{person}{VisionAware}.}
  \bibinfo{year}{[n.d.]}\natexlab{}.
\newblock \bibinfo{title}{Safe Cooking Techniques for Cooks Who Are Blind or
  Have Low Vision - VisionAware}.
\newblock
  \bibinfo{howpublished}{\url{https://visionaware.org/everyday-living/essential-skills/cooking/safe-cooking-techniques/}}.
\newblock
\newblock
\shownote{(Accessed on 04/05/2021).}


\bibitem[\protect\citeauthoryear{Wiberg}{Wiberg}{2021}]%
        {BeMyEyes54:online}
\bibfield{author}{\bibinfo{person}{Hans~Jørgen Wiberg}.}
  \bibinfo{year}{2021}\natexlab{}.
\newblock \bibinfo{title}{Be My Eyes - See the world together}.
\newblock \bibinfo{howpublished}{\url{https://www.bemyeyes.com/}}.
\newblock
\newblock
\shownote{(Accessed on 04/10/2021).}


\bibitem[\protect\citeauthoryear{Wikipedia}{Wikipedia}{2021}]%
        {Christin65:online}
\bibfield{author}{\bibinfo{person}{Wikipedia}.}
  \bibinfo{year}{2021}\natexlab{}.
\newblock \bibinfo{title}{Christine Ha - Wikipedia}.
\newblock
  \bibinfo{howpublished}{\url{https://en.wikipedia.org/wiki/Christine_Ha}}.
\newblock
\newblock
\shownote{(Accessed on 03/14/2021).}


\bibitem[\protect\citeauthoryear{Ye, Malu, Oh, and Findlater}{Ye
  et~al\mbox{.}}{2014}]%
        {ye2014current}
\bibfield{author}{\bibinfo{person}{Hanlu Ye}, \bibinfo{person}{Meethu Malu},
  \bibinfo{person}{Uran Oh}, {and} \bibinfo{person}{Leah Findlater}.}
  \bibinfo{year}{2014}\natexlab{}.
\newblock \showarticletitle{Current and future mobile and wearable device use
  by people with visual impairments}. In \bibinfo{booktitle}{\emph{Proceedings
  of the SIGCHI Conference on Human Factors in Computing Systems}}.
  \bibinfo{pages}{3123--3132}.
\newblock


\bibitem[\protect\citeauthoryear{Zhou, Li, and Zhou}{Zhou
  et~al\mbox{.}}{2017}]%
        {zhou2017iot}
\bibfield{author}{\bibinfo{person}{Mingyong Zhou}, \bibinfo{person}{Wenyan Li},
  {and} \bibinfo{person}{Bo Zhou}.} \bibinfo{year}{2017}\natexlab{}.
\newblock \showarticletitle{An IOT system design for blind}. In
  \bibinfo{booktitle}{\emph{2017 14th Web Information Systems and Applications
  Conference (WISA)}}. IEEE, \bibinfo{pages}{90--92}.
\newblock


\end{thebibliography}


\end{document}